# Universal scattering behavior of co-assembled nanoparticle-polymer clusters

**J. Fresnais[1], J.-F. Berret[1,®], L. Qi[2], J.-P. Chapel[2], J.-C. Castaing[2], O. Sandre[3], B. Frka-Petesic[3], R. Perzynski[3], J. Oberdisse[4] and F. Cousin[5]**

*(1) : Laboratoire Matière et Systèmes Complexes (MSC), UMR 7057 CNRS & Université Paris Diderot,*
*Bâtiment Condorcet, 10 rue Alice Domon et Léonie Duquet, 75205 Paris, France*
*(2) : Complex Fluid Laboratory, FRE CNRS/Rhodia 3084, Rhodia North-America, R&D Headquarters CRTB,*
*350 George Patterson Blvd., Bristol, PA 19007 USA*
*(3) : UPMC Univ Paris 06 - Laboratoire Liquides Ioniques et Interfaces Chargées, UMR 7612 CNRS,*
*4 place Jussieu-case 63, F-75252 Paris Cedex 05 France*
*(4) : Laboratoire des Colloïdes, Verres et Nanomatériaux, UMR 5587 CNRS, Université Montpellier II, 34095, Montpellier, France*
*(5) : Laboratoire Léon Brillouin, UMR CEA-CNRS 12, CEA- Saclay, 91191 Gif-sur-Yvette, France*

## Abstract

Water-soluble clusters made from 7 nm inorganic nanoparticles have been investigated by small-angle neutron scattering. The internal structure factor of the clusters was derived and exhibited a universal behavior as evidenced by a correlation hole at intermediate wave-vectors. Reverse Monte-Carlo calculations were performed to adjust the data and provided an accurate description of the clusters in terms of interparticle distance and volume fraction. Additional parameters influencing the microstructure were also investigated, including the nature and thickness of the nanoparticle adlayer.

The synthesis of inorganic nanocrystals using soft chemistry routes has recently opened new strategies for the organization and structure of matter at the nanometer range. Inorganic nanoparticles made from noble metals (Au, Pt, Rh), from oxides (iron, cerium, titanium) or from semiconductors (CdSe/ZnS) and dispersed in organic or aqueous solvents are characterized by sizes between 1 and 100 nm. The existence of a crystalline structure confined at nanometric length scales confers to the particles remarkable physical properties in optics and in magnetism. It is anticipated that these particles will play an increasingly important role in the development of nanoscience. In the present work, we have studied iron and cerium oxide nanoparticles ($\gamma$-$Fe_2O_3$ and $CeO_2$), both with sizes 7 nm and dispersed in water. In terms of applications, $\gamma$-$Fe_2O_3$ particles are utilized in magnetic resonance imaging as contrast agents and in therapy as the support for hyperthermia[1], whereas cerium oxide remain an essential ingredient in catalysis, coating and biomedecine[2-4].

Particulates of larger sizes *i.e.* in the range 50 nm – 1 µm may be required in applications (e.g. chemical-mechanical polishing[5]), but yet remain difficult to generate by soft chemistry routes. Several issues become critical when the particle sizes increase, including colloidal stability, sedimentation and loss or reduction of physical properties. Because of their large mass densities, metal oxide particles in 100 nm range exhibit for instance sedimentation kinetics, which often results in uncontrolled aggregation. Magnetic nanoparticles are also known for their remarkable size-related properties. Above 30 nm however, a reduction of the remanent magnetization due to multidomain structure is observed, hindering their uses in applications[1]. In order to circumvent such limitations, self- or co-assembly strategies involving nanoparticles have been suggested[6-17]. Best examples are the "brick and mortar" approach developed by Boal et al.[6,8], various complexation schemes using nanoparticles and ion-containing polymers[13-17] or spray-drying nanocomposite pow-ders[18]. In such instances, nanoparticles were arranged into clusters of different dimensions and morphologies, including spheres[6,10,13,16,18], wires[14] and filaments[9,17], with aggregation numbers ranging from few units to several thousands. Although aggregates of nanoparticles are nowadays regarded as a promising type of colloids, their microstructures have been relatively unexplored, in particular with respect to small-angle scattering techniques. Using small-angle neutron scattering (SANS), we have identified the ubiquitous signatures of nanoparticle clusters, and in particular we have shown evidences of a correlation hole in the structure factor of the aggregates. Reverse Monte-Carlo simulations were performed to fit the SANS data, and provided the cluster microstructure with a high degree of accuracy. From the various parameters that may impact the cluster microstructure, the organic adlayer around the particles was found to be the most significant as compared to the nature of the particles or the molecular weight of the complexing polyelectro-lytes.

The formation of nanoparticle clusters was obtained by electrostatic complexation between anionically charged nanoparticles and cationic-neutral block copolymers. As for polymers, poly(trimethylammonium ethylacry-

late)-b-poly(acrylamide) with different molecular weights were used. The two polymers investigated, hereafter noted PTEA$_{5K}$-*b*-PAM$_{30K}$ and PTEA$_{11K}$-*b*-PAM$_{30K}$ were displaying an excellent dispersability in aqueous solvents19 (hydrodynamic diameter $D_H$ = 11 nm).

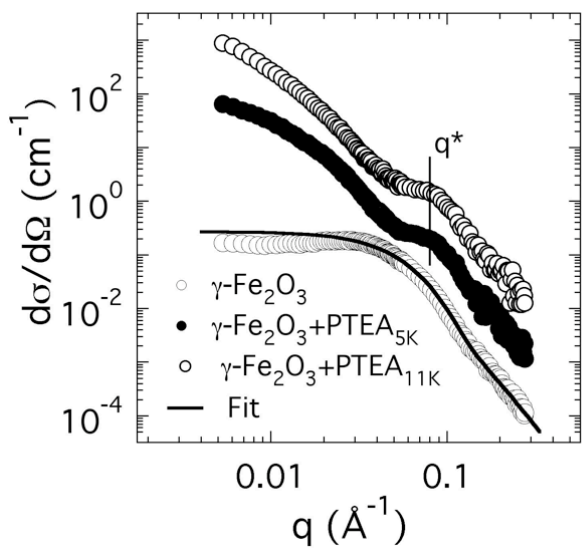

**Figure 1** : Scattering cross-sections for citrate-coated iron oxide nanoparticles, nanoparticles complexed with PTEA$_{5K}$-*b*-PAM$_{30K}$ and with PTEA$_{11K}$-*b*-PAM$_{30K}$ block copolymers. The total concentrations in actives were 6.3, 2 and 2 wt. %, respectively. The experimental intensities were shifted by coefficients 1/80, 1 and 6 for sake of clarity

The nanoparticles dispersions were characterized by cryo-TEM and light scattering, and for superparamagnetic $\gamma$-Fe$_2$O$_3$ by vibrating sample magnetometry[20]. The size distributions were found to be well-accounted for by log-normal functions, with median diameter D around 7 nm and a polydispersity s of 0.2 (Table I). In the conditions at which the complexation was realized (neutral pH and no added salt), the particles were stabilized by electrostatic interactions mediated either by charged citrate ligands or by short poly(acrylic acid) oligomers of molecular weight 2000 g mol$^{-1}$. ζ-potential measurements (Zetasizer Nano ZS, Malvern Instrument) were performed and have shown that the particles were negatively charged[15]. The hydrodynamic size of the PAA$_{2K}$-coated nanoparticles was found to be 4 nm larger than the hydrodynamic diameter of the uncoated ones, a result that was interpreted as arising from the PAA$_{2K}$ brush around the nanoparticles[20].
Hybrid clusters were prepared by simple mixing of stock nanoparticle and copolymer solutions. The formation of the mixed colloids is qualitatively understood as a nucleation and growth process of a microphase made from the oppositely charged constituents. The growth is arrested at a size which is typically fixed by the length of the neutral block[19,15]. The relative amount of each species was monitored by the charge ratio, which was fixed to unity[15].
Small-angle neutron scattering was carried out at the Laboratoire Léon Brillouin (Saclay, France) on the PAXY beam line. For SANS, polymer-nanoparticle hybrids were prepared in H$_2$O for contrast reasons : the scattering length densities for CeO$_2$ and for $\gamma$-Fe$_2$O$_3$ were $5.0 \times 10^{10}$ cm$^{-2}$ and $7.2 \times 10^{10}$ cm$^{-2}$, respectively. Since the density for acrylamide amounts to $\rho_N(AM)$ = + $1.86 \times 10^{10}$ cm$^{-2}$, we anticipate in H$_2$O ($\rho_N$ = -$0.56 \times 10^{10}$ cm$^{-2}$) the cores of the clusters will contribute predominantly to the overall scattering cross-section19. The data collected at 1.35 and 6.75 meters cover a range in wave-vector : $5 \times 10^{-3}$ Å$^{-1}$ to 0.28 Å$^{-1}$, with an incident wavelength of 6 Å and a wave-vector resolution $\Delta q/q$ of 10 %.

Fig. 1 compares the scattering cross-sections d$\sigma$(q)/d$\Omega$ for citrate-coated iron oxide particles and of clusters obtained through complexation with PTEA$_{5K}$-*b*-PAM$_{30K}$ and PTEA$_{11K}$-*b*-PAM$_{30K}$ block copolymers. For citrate-coated $\gamma$-Fe$_2$O$_3$ particles, the intensity was adjusted using a model of polydisperse spheres (see continuous line in Fig. 1), and a log-normal size distribution with median diameter D = 5.5 nm and polydispersity s = 0.28. These latter data were in good agreement with those obtained from cryo-TEM or vibrating sample magnetometry on the same system20.

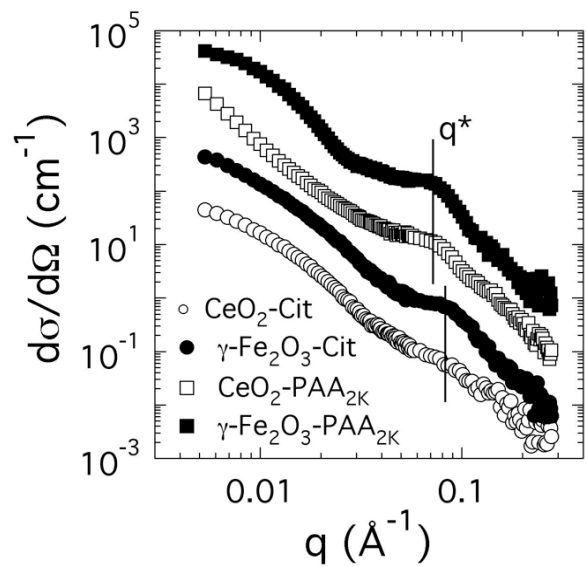

**Figure 2** : SANS cross-sections of nanoparticle-copolymer clusters comparing citrate and PAA$_{2K}$-coated CeO$_2$ and $\gamma$-Fe$_2$O$_3$. The total concentrations were c = 5.8, 2, 11 and 1.7 wt. % for the clusters made with CeO$_2$-Cit, $\gamma$-Fe$_2$O$_3$-Cit, CeO$_2$-PAA$_{2K}$ and for $\gamma$-Fe$_2$O$_3$-PAA$_{2K}$ respectively. The intensities were shifted upwards by coefficients 1, 3, 2 and 500.

By contrast, the clusters exhibited different scattering features, which are *i)* a strong forward scattering (q → 0), *ii)* a shoulder in the intermediate q-range centered at wave-vector q* = 0.084 ± 0.003 Å$^{-1}$ (similar for both polymers) and *iii)* a q$^{-4}$-decrease of the intensity at high wave-vectors. These three features were identical to those of colloidal complexes obtained with surfactant micelles[19]. By analogy with the work on micelles, the shoulder observed at q* for γ-Fe$_2$O$_3$-Cit/PTEA$_{5K}$-*b*-PAM$_{30K}$ and γ-Fe$_2$O$_3$-Cit/PTEA$_{11K}$-*b*-PAM$_{30K}$ was here interpreted as the first order peak of the particle-particle structure factor. Assuming a simple scaling between the interparticle distance d and q*, one gets d = 2π/q* = 7.5 ± 0.2 nm for the two systems. Note that the polydispersity was more important for the oxides than for surfactant micelles and as a result, the scattering cross-sections were smeared out. Fig. 2 exhibits the scattering cross-sections for complexes obtained using CeO$_2$ and γ-Fe$_2$O$_3$ nanoparticles coated either with small citrate ions or with larger poly(acrylic acid) oligomers. The same copolymer, PTEA$_{11K}$-*b*-PAM$_{30K}$ was used in the four mixed systems. Form factors for bare and coated CeO$_2$ particles were reported earlier[21].

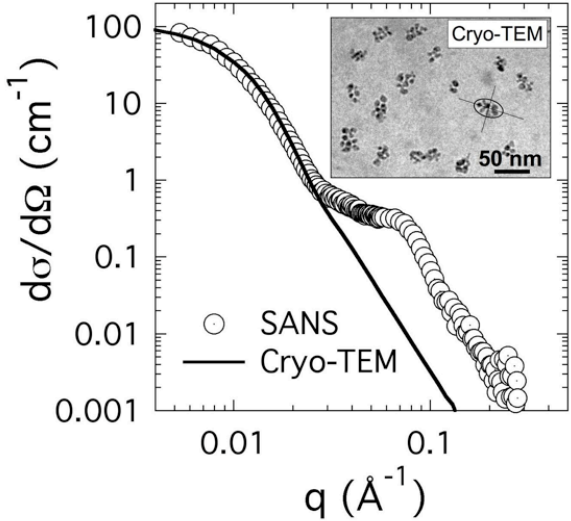

**Figure 3** : Same data as in Fig. 2 for γ-Fe$_2$O$_3$-PAA$_{2K}$/PTEA$_{11K}$-*b*-PAM$_{30K}$, together with a calculation of the form factor of polydisperse ellipsoids. Inset : Cryo-TEM image of the clusters. Transmission electron cryo-microscopy was performed at liquid nitrogen temperature using a JEOL 1200 EX microscope operating at 120 kV with a ×40000 magnification[15]. Due to the low electronic contrast between the copolymers and water, the presence of acrylamide moieties around the aggregates could not be revealed here.

The scattering features revealed in Fig. 1 were recovered. Note the shift of the nanoparticle structure factor peak to lower wave-vectors for the PAA$_{2K}$-coated particles (q* = 0.073 Å$^{-1}$, corresponding to d = 8.6 ± 0.2 nm), as compared to the citrate-coated particles.

Fig. 3 displays the neutron scattering intensity and cryo-microcopy characterization of a same γ-Fe$_2$O$_3$-PAA$_{2K}$/PTEA$_{11K}$-*b*-PAM$_{30K}$ sample. The scattering data are those of Fig. 2. In the low q-region, a Guinier analysis provided a gyration radius R$_G$ = 19.2 nm. The photograph in the inset confirms the existence of dispersed aggregates. From an image analysis of 210 aggregates, these colloids were at best described by slightly elongated ellipsoids, with an average major axis 31 nm and aspect ratio 1.5. From the ellipsoid distribution, the scattering cross-section was then calculated and compared to the SANS data (continuous line in Fig. 3). The agreement between the two techniques is remarkable in the low-q region. Above 3×10$^{-2}$ Å$^{-1}$ however, the calculated curve under-estimated the SANS data by a factor 20, the reason for this being related to the heterogeneous microstructure of the clusters.

In order to interpret the neutron scattering data of Fig. 2, we recall the expression of the scattering form factor for an N-aggregate built up of identical spheres[19]. The aggregation number N >> 1 was first related to the cluster and particle diameters D$_C$ and D respectively, and to the volume fraction φ$_C$ inside the cluster through: N = φ$_C$(D$_C$/D)$^3$. For a dispersion containing n(φ) small spheres per unit volume arranged into N-clusters, the scattering cross-section was expressed as[18,19,22]:

$$\frac{d\sigma_C}{d\Omega}(q,N,\phi) = n(\phi)\Delta\rho^2 V^2 F(q,R) S_C(q,N) \qquad (1)$$

where $S_C(q,N) = 1 + \frac{2}{N}\left\langle \sum_{i=1}^{N-1}\sum_{j=i+1}^{N}\frac{\sin qr_{ij}}{qr_{ij}} \right\rangle$.

Δρ denotes the scattering contrast and F(q,R) the form of individual spheres. In Eq. 1, $r_{ij} = |\mathbf{r_i} - \mathbf{r_j}|$ is the distance between the center of mass of the spheres. The brackets in the definition of S$_C$(q,N) mean the averaging over all positional configurations of the nanoparticles inside the core.

An important result of Eq. 1 is that the scattering expressed as a product of the form factor of the elementary particles and the function S$_C$(q,N), also called the structure factor of a N-cluster. The asymptotic limits of S$_C$(q,N) are also remarkable, since S$_C$(q,N) goes to N as q → 0, and to 1 as q → ∞. With this in mind, the experimental data of Fig. 2 were re-examined. The cluster intensities have been divided by the cross-sections of the individual particles extra-polated at the same total volume fraction, and plotted as a function of the product $q\langle D\rangle$ where $\langle D\rangle$ is the average nanoparticle diameter. Fig. 4 shows the structure factors for the 4 types of clusters put under scrutiny. The main features for the structure factor are a strong increase at low wave-vectors, a correlation hole in the intermediate q-range and an asymptotic behavior toward unity in the high-q limit.

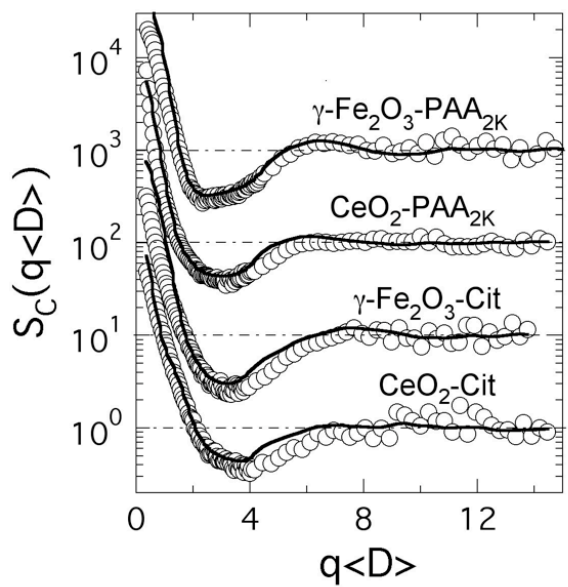

**Figure 4** : Internal structure factors of clusters as a function of the product $q\langle D\rangle$, where $\langle D\rangle$ is the average nanoparticle diameter. Continuous lines are from Reverse Monte Carlo simulations. There, the single particles were described as polydisperse and obeying the size distributions found experimentally.

Characterized by values less than unity, the correlation hole is indicative of a structure dominated by two separated length scales, which are here the size of the particles and that of the clusters. If these two dimensions are away from each other by a factor 3, meaning $D_C/D > 3$, the scattering of the clusters will be below that of the particles, yielding a dip in the structure factor[23]. Concerning the low q-increase, note that except for $\gamma$-Fe$_2$O$_3$-PAA$_{2K}$/PTEA$_{11K}$-$b$-PAM$_{30K}$ the clusters were too large to display an asymptotic Guinier behavior within the present q-window.

We turn now to the quantitative analysis of the neutron data using Reverse Monte Carlo technique[22] (RMC). The RMC method consisted in generating representative aggregates by moving elements of the aggregate in a random way and in calculating the corresponding structure factor at each step. The comparison with the experimental provided a criterion whether the Monte Carlo step was accepted or not. For each scattering curve, simulations were repeated at least 10 times and the generated aggregates were checked to be consistent with each others. The simulated $S_C(q)$'s in Fig. 4 are averages over these different runs. In order to account for the coating of the particles, the size distributions were shifted by a quantity noted h. h represented thus the thickness of the organic adlayer around the particles. The RMC simulations were performed by varying h and N. For the citrate coated cerium and iron particles, we have found h = 0, corresponding to volume fractions around 0.30, whereas for the PAA$_{2K}$-coated particles, we obtained h = 0.7 nm, with a volume fraction around 0.20.

| clusters made from | D (s) (nm) | d$_{RMC}$ (nm) | φ | N |
|---|---|---|---|---|
| CeO$_2$-Cit | 6.9 (0.15) | 7.1 | 0.30 | 150 |
| $\gamma$-Fe$_2$O$_3$-Cit | 6.3 (0.23) | 6.55 | 0.38 | 100 |
| CeO$_2$-PAA$_{2K}$ | 6.9 (0.15) | 8.5 | 0.21 | 100 |
| $\gamma$-Fe$_2$O$_3$-PAA$_{2K}$ | 7.1 (0.26) | 8.7 | 0.23 | 30-60 |

**Table I** : List of parameters used for the fitting of the structure factors in Fig. 4. The interparticle distance d$_{RMC}$, the volume fraction φ and the aggregation number N were estimated from Monte Carlo simulations. All clusters were made using PTEA$_{11K}$-$b$-PAM$_{30K}$ block copolymers.

For the four systems, the simulations were able to reproduce the $S_C(q)$'s correctly, and in particular the dip observed at intermediate wave-vectors. As anticipated, the mean distances between the particles dRMC were found to be slightly larger than the bare diameter, by 0.2 nm for citrate and by 1.6 nm for PAA$_{2K}$ (Table I). These values are more accurate than the estimates obtained from the position of the structure peaks at q*. For magnetic nanoparticles, interparticle distance and internal volume fraction are important parameters, since they determine the magnetic dipolar energy and the susceptibility of this novel magnetic material. In conclusion, we have shown that SANS allows to identify the structure of co-assembled clusters through the presence of a universal scattering signature independent of the particle, organic adlayer and complexing agent. The occurrence of such correlation hole in the internal structure factor of the aggregates is indicative of two well separated hierarchical length scales in these systems: the size of the particles and that of the clusters. Interpretations were performed in combination with reverse Monte Carlo simulations.